\documentclass[aps,pre,showpacs,twocolumn,groupedaddress]{revtex4}

\usepackage{amsmath,amssymb}
\usepackage{graphicx}
\usepackage{psfrag}

\newcommand{\lsbar}{\bar l_s}
\newcommand{\boldface}{}

\begin{document}


\title{The Role of Architecture in the Elastic Response of Semi-flexible Polymer
  and Fiber Networks} \author{Claus Heussinger} \author{Erwin Frey}

\affiliation{Arnold Sommerfeld Center for Theoretical Physics and
  CeNS, Department of Physics, Ludwig-Maximilians-Universit\"at
  M\"unchen, Theresienstrasse 37, D-80333 M\"unchen, Germany}

\affiliation{Hahn-Meitner-Institut, Glienicker Strasse 100, D-14109
  Berlin, Germany}

\begin{abstract}
  We study the elasticity of cross-linked networks of thermally fluctuating
  stiff polymers. As compared to their purely mechanical counterparts, it is
  shown that these thermal networks have a qualitatively different elastic
  response. By accounting for the entropic origin of the single-polymer
  elasticity, the networks acquire a strong susceptibility to polydispersity and
  structural randomness that is completely absent in athermal models.  In
  extensive numerical studies we systematically vary the architecture of the
  networks and identify a wealth of phenomena that clearly show the strong
  dependence of the emergent macroscopic moduli on the underlying mesoscopic
  network structure. In particular, we highlight the importance of the polymer
  length that to a large extent controls the elastic response of the network,
  surprisingly, even in parameter regions where it does not enter the
  macroscopic moduli explicitly. Understanding these subtle effects is only
  possible by going beyond the conventional approach that considers the response
  of {\it typical polymer segments} only.  Instead, we propose to describe the
  elasticity in terms of a {\it typical polymer filament} and the spatial
  distribution of cross-links along its backbone. We provide theoretical scaling
  arguments to relate the observed macroscopic elasticity to the physical
  mechanisms on the microscopic and the mesoscopic scale.
\end{abstract}

 \date{\today}

\maketitle

\section{Introduction}

Classical elasticity is a continuum theory that deals with the large scale
deformation properties of solid systems. It relates stresses and strains by
introducing a host of phenomenological parameters, e.g. shear and bulk modulus
for isotropic media, that characterize the elastic properties on wave lengths
large compared with any other material length scale~\cite{landau7}. Biological
systems like the cell or sub-cellular organelles are often characterized by a
highly heterogeneous structure with a multitude of hierarchical levels of
organization~\cite{alb94}. Due to these large scale inhomogeneities that may
extend up to the scale of the system size, the applicability of elasticity
theory on smaller length scales has to be critically examined. In particular,
the actual deformations in the system are expected to relate to the externally
applied stresses in a non-trivial way that crucially depends on the specific
structural details.

To shed some light on the relevance of structure to the effective elasticity
this article deals with the calculation of elastic constants in networks of
semi-flexible polymers. In eukaryotic cells these networks assemble to form the
cytoskeleton that plays a central role in many cellular functions such as
locomotion, adhesion or cell division. From the point of view of structure
already a one-component isotropic solution of semi-flexible polymers represents
an interesting model-system being studied for many years~\cite{hin98,xu98,cla}.
One of the main quantities of interest is the plateau value of the shear modulus
found at intermediate timescales where single polymer bending fluctuations are
equilibrated, yet center of mass motion is negligible.  The generally accepted
theory for the concentration dependence of the plateau modulus is based on the
free energy change of confining a polymer to a
tube~\cite{odj83,hel85,hin98,isa96}, the diameter of which is a consequence of
the structural organization of the tubes in the form of a random assembly of
cylinders~\cite{phi96}. Even though this is well known for more than a decade,
computer simulations to study the geometrical as well as elastic properties in
this fibrous architecture have only recently been realized~\cite{rod05,willi03}.

Upon the addition of cross-linking agents or other regulating proteins one can
induce structural changes to modify the network architecture in many
ways~\cite{tem96,pelletier03,tha06,limozin02,shi04}. There have been attempts to
describe the phase-diagram of these systems~\cite{bor05,zil03}, the detailed
mechanisms that lead to a particular structure, however, are far from being
understood.  In general, there will be a complex interplay of polymer kinetics,
thermal fluctuations and chemical as well as mechanical properties of the
polymers and the cross-linking agents yielding a particular architecture
relevant for a given physical situation.

A complementary approach to describe cross-linked networks is to neglect these
intricate ``dynamic'' aspects of the network, and to concentrate on a ``static''
architecture and its effect on the macroscopic
elasticity~\cite{frey98,wil03,hea03a,hea03c,onck05,heu06a}. In the structural
engineering community, for example, it is of tantamount importance to analyze
the architecture of structures made of beams or trusses. A common way to take
advantage of the reduced weight compared to the bulk material without suffering
from a loss of stiffness is a triangulation of the basic cells. This eliminates
the soft bending modes of the beams and makes it possible to construct huge
cantilever bridges like that over the Firth of Forth in Scotland or towers like
Eiffel's tower in Paris. Since the rigidity of these structures is not due to
the individual beam but to a non-local back-coupling effect induced by the
architecture of the network, the triangulation is therefore one example on how
cooperativity among the building blocks may be possible.

\begin{figure}[tb!]
\begin{center}
  \includegraphics[width=0.9\columnwidth]{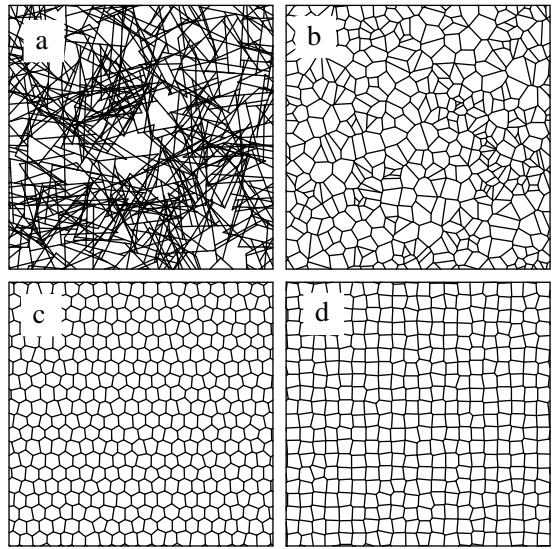}
\end{center}
\caption{\label{fig:vorNet} Illustration of the different architectures of (a)
  fibrous and (b-d) cellular materials in two dimensions. While (a) and (b) are
  random structures generated by Poisson point processes, (c) and (d) are quite
  regular networks based on honeycomb and square lattices, respectively.}
\end{figure}

To address this question of cooperativity in the context of the elasticity of
cross-linked stiff polymer networks we will concentrate in the following on two
generic structures, {\it cellular} and {\it fibrous} networks, that may serve as
reference systems for the classification of real polymer networks. While
cellular structures may be characterized by the amount of randomness in size and
type of their unit cells (see Fig.\ref{fig:vorNet}b-d), fibrous networks have a
hierarchical structure, where smaller cells are generated within lager cells
within even larger cells (Fig.\ref{fig:vorNet}a). This is a consequence of the
presence of the additional mesoscopic scale of the fiber length. As we will see,
this length-scale is ultimately responsible for the intricate scaling properties
of the elasticity of fibrous polymer networks. The goal of this article is to
identify these mechanisms, that couple the particular network structure to the
properties of the individual polymers and effectuate the macroscopic elasticity
of the system.

In contrast to the purely mechanical systems relevant for engineering
applications~\cite{ast00,wil03,hea03a,hea03c}, the systems we would like to
study are immersed in a thermal environment. This implies that in addition to
the usual enthalpic polymer elasticity also entropic effects have to be
accounted for. We have published a brief account of this study
recently~\cite{heu06a}.  It will turn out that by accounting for the entropic
origin of the single-polymer elasticity, the networks acquire a strong
susceptibility to polydispersity and structural randomness that is completely
absent in athermal models.

The article is organized as follows. In Sect.\ref{sec:model-definition} we
motivate our modeling approach of thermally fluctuating networks of stiff
polymers. This will lead us to the definition of effective elastic properties of
the ``polymer segments'' that constitute the elementary building blocks of the
network. In Sect.\ref{sec:cell-arch} and \ref{sec:fibrous-architecture} these
polymer segments are assembled into cellular and fibrous networks, respectively.
The macroscopic elastic constants of these structures are calculated and related
to the particular architectural features. Finally, in Sect.\ref{sec:conclusion}
we present our main conclusions and hint at implications for experiments.

\section{Model Definition}\label{sec:model-definition}

\begin{figure}[tb!]
\begin{center}
  \includegraphics[width=0.8\columnwidth]{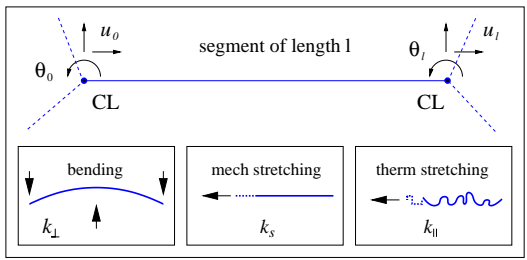}
  \end{center}
  \caption{\label{fig:modeling} Illustration of a polymer segment of length $l$
    and its connection to the network (dashed lines) at the cross-links (CL).
    The three degrees of freedom at each cross-link are denoted by $\bf u$ and
    $\theta$, respectively. Identification of the three possible modes of
    deformation and their stiffnesses $k_\perp$, $k_s$ and $k_\parallel$ as
    defined in the text.}
\end{figure}

To study the elastic properties of thermally fluctuating cross-linked stiff
polymer networks we calculate numerically the low frequency shear modulus.
Assuming a time-scale separation between the fast bending fluctuations of the
single polymer and their very slow center of mass motion, we adopt a description
of the system in the spirit of a Born-Oppenheimer approximation. This neglects
entropic contributions from the ``slow variables'', the cross-link positions,
while assuming the ``fast'' polymer degrees of freedom to be equilibrated at all
times.  {\boldface Macroscopic quantities will then depend parametrically on the set
  of cross-link variables. A macroscopic shear strain $\gamma$ constrains the
  cross-links at the boundaries, while those in the bulk are moving freely to
  minimize the elastic energy $E$. The shear modulus is defined as its second
  derivative with respect to the shear strain, $G=V^{-1}\partial^2E_{\rm
    min}/\partial\gamma^2$, where $V$ is the system volume.

  By keeping the positions of the cross-links fixed, the energy can be written as
  a sum
  \begin{equation}\label{eq:internEn}
    E = \sum_\alpha e(\boldsymbol{\delta x}^\alpha)\,,
  \end{equation}
  over contributions from individual polymer segments $\alpha$, each of which
  connects a given pair of cross-links (see Fig.\ref{fig:modeling}). The single
  segment energy $e$ depends on the generalized ``displacement-vector''
  $\boldsymbol{\delta x}^\alpha$, which incorporates the degrees of freedom,
  displacements $\boldsymbol{u}$ and rotations $\boldsymbol{\theta}$, of the two
  cross-links pertaining to the segment.

  In the numerical section we focus on two-dimensional systems such that a
  vector $\boldsymbol{\delta x}_{\rm 2d} = (\boldsymbol{u}_0, \theta_0,
  \boldsymbol{u}_l, \theta_l)$ has six components. Those are in-plane
  displacements $\mathbf{u}_{0,l}$ and z-axis rotations $\theta_{0,l}$, at both
  ends $0,l$ of the segment with length $l$ (see Fig.\ref{fig:modeling}).}
Note, that the additional variable of cross-link rotation sets our system apart
from bond-bending and related models~\cite{sahimi} where only translational
degrees of freedom are accounted for. As a consequence one also has to account
for the presence of torques as the conjugate variable to rotations.

To leading order, in linear elasticity, the single segment quantity $e$ is a
quadratic function of its coordinates
\begin{equation}\label{eq:harmonEn}
  e(\mathbf{x}) = \frac{1}{2}\mathbf{x^TK x}\,,
\end{equation}
which defines the ``stiffness matrix'' $\mathbf{K}$ (spring constants) of the
polymer strand.

{\boldface In models of classical beams with cross-section radius $r$ the matrix
  elements are well established and relate to the two deformation modes of
  stretching (s) and bending ($\perp$), respectively. While the former is
  characterized by the Young's modulus $E$ of the material, the latter depends
  on the bending stiffness $\kappa=E\pi r^4/4$, here taken for circular
  cross-sections. To calculate the bending response, standard Euler-Bernoulli
  beam theory~\cite{landau7} is used.

  While we refer to Appendix~\ref{sec:stiffness-matrix} for a derivation of the
  complete matrix, it turns out that the response of a beam of length $l$ is
  sufficiently characterized already by two elements of $\mathbf{K}$,
  \begin{equation}\label{eq:kbeam}
    k_s(l)=4\kappa/lr^2\,, \qquad   k_\perp(l)=3\kappa/l^3\,,
  \end{equation}
  relating to either deformation mode. Due to their small aspect ratios $r/l \ll
  1$ slender rods are highly anisotropic and much softer in bending than in
  stretching, $k_\perp/k_s \propto (r/l)^2$. In this approximation the two
  deformation modes are decoupled such that, for example, pre-stretching does
  not influence the bending response.  Therefore, Euler-buckling cannot be
  accounted for. }

Here, we consider thermally fluctuating stiff polymers immersed in a heat bath
of solvent molecules. In these systems, the effects of temperature on the
elastic properties of the polymer can be quantified by defining the persistence
length $l_p$ as the ratio of bending stiffness to thermal energy
$l_p=\kappa/k_BT$. With this definition we have, in addition to the enthalpic
stiffness of the classical beam, an entropic contribution
\begin{equation}\label{eq:k_entr}
  k_{\parallel} (l) = \zeta \, \kappa \, \frac{l_p}{l^4}\,,
\end{equation} 
to the polymer's stretching compliance that can be calculated within the
wormlike chain model~\cite{mac95,kro96}. The prefactor $\zeta$ depends on the
specific boundary conditions chosen at the ends of the polymer segment. Its
value can be absorbed in the persistence length, and therefore only
quantitatively affects the results. To avoid a large numerical offset with
respect to Eq.(\ref{eq:kbeam}), we have chosen $\zeta=6$, which corresponds to a
boundary condition with one end clamped~\cite{kro96}. Having two longitudinal
deformation modes $k_s$ and $k_\parallel$ the effective stretching stiffness is
equivalent to a serial connection
\begin{equation}\label{eq:eff_k_stretch}
  k_{\rm eff}^{-1}=k_s^{-1}+k_\parallel^{-1}\,.
\end{equation}

{\boldface Thus, the elastic properties of the polymer segments are given by the
  classical theory of beam bending supplemented by a generalized stretching
  stiffness, that also includes entropic effects. While the stiffness matrix has
  only been set up for the two-dimensional problem, the governing entries in
  three dimensions will still be the same Eqs.(\ref{eq:kbeam}) and
  (\ref{eq:k_entr}).}

As one can infer from Eqs.(\ref{eq:kbeam}) and (\ref{eq:k_entr}), at a given
temperature $T$ there are two length scales characterizing the material
properties of the polymers, the radius $r$ and the persistence length $l_p$.
Typical biological polymers are characterized by a ratio $R=l_p/r \gg 1$.
F-actin, for example, a key component of the cytoskeleton has $R=O(10^4)$
($r\approx 5{\rm nm}\,,\,\,l_p\approx 17{\rm \mu m}$), while microtubules, most
important for cell-division and intra-cellular transport, have an even larger
$R=O(10^6)$.  For specificity, we require in the following a constant
$R=1.5\cdot 10^4$, the precise value, however, is irrelevant if one is
interested only in the thermal response where the radius does not enter and
$k_s\to\infty$~\footnote{This limit corresponds to taking $r\to 0$ and the
  Youngs-modulus $E\to\infty$ in such a way that the bending stiffness
  $\kappa\sim Er^4$ stays constant. As a consequence the stretching stiffness
  diverges, $k_s\sim Er^2\to\infty$, and the beam becomes inextensible.}.
Occasionally, we will perform this limit to highlight features that are
independent of the mechanical stretching response.  On the other hand, the
location of the cross-over point, where the mechanical stretching becomes
relevant, does indeed depend on the choice of $R$. By definition, it determines
the relative magnitude of the two stretching compliances $k_s/k_\parallel\simeq
R^2(l/l_p)^3$.

\begin{figure}[tb!]
\begin{center}
  \includegraphics[width=0.8\columnwidth]{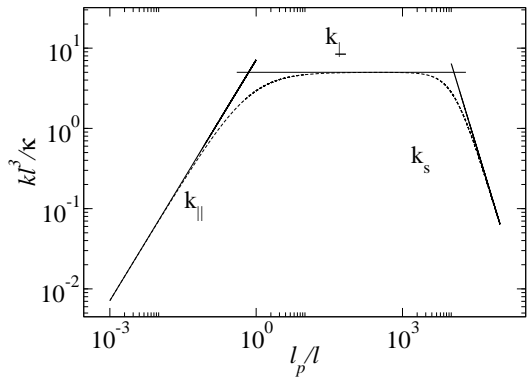}
  \end{center}
  \caption{\label{fig:forceconstants} Dependence of the three spring constants
    $k_\parallel$, $k_\perp$ and $k_s$ on persistence length $l_p/l$. The
    dashed line corresponds to a hypothetical spring with the three deformation
    modes connected in series.}
\end{figure}

The dependence of the three force constants $k_\perp$, $k_s$ and $k_\parallel$,
Eqs.(\ref{eq:kbeam}) and (\ref{eq:k_entr}), on the ratio of persistence length
to segment length $l_p/l$ is illustrated in Fig.\ref{fig:forceconstants}. One
can clearly distinguish three regimes, in each of which one of the spring
constants is by far smaller than the remaining two. The dashed line corresponds
to a hypothetical spring where the deformation modes are coupled in series
$k^{-1} = k_\perp^{-1}+k_s^{-1}+k_\parallel^{-1}$. If the segment length $l$ was
representative for the network under consideration, that is the network was
characterized by only small polydispersity, then we would expect the macroscopic
modulus to be well approximated by the microscopic single segment behavior
considered here.  We will later refer to this behavior as the ``affine model''.
It will be shown to be valid only in regular cellular structures.

This completes the specification on the microscopic scale of the elastic
properties of the single polymer segments. We now proceed to assemble the
segments into networks of varying architecture to identify the physical
principles which determine the elastic response on the macroscopic scale.

To determine the elastic shear modulus, we apply a small shear strain of
$\gamma=0.01$ to stay in the regime of linear elasticity and use periodic
boundary conditions on all four sides of the simulation box. The numerical
procedure is performed with the commercially available finite element solver
MSC.MARC. The results will be complemented by scaling arguments.

We will find that in {\it regular cellular architectures}, to be discussed next,
macroscopic elasticity can trivially be explained by the microscopic
constitutive laws given in terms of the stiffness matrix $\bf{K}$. In
sufficiently \emph{random cellular systems}, however, this picture is changed.
The macroscopic response takes up nontrivial features that cannot be explained
by single polymer elasticity. In \emph{fibrous architectures}, subject of
Sect.\ref{sec:fibrous-architecture}, we will find this anomalous elasticity
again but in more striking form.

\section{Cellular Architecture}\label{sec:cell-arch}

A cellular structure is most conveniently constructed from a Voronoi tessellation
of a distribution of points which may either be chosen regularly or by some
random process~\cite{wea01}. With each point we associate a Voronoi cell that is
defined to enclose that region in space which is closer to the given point than
to any of its neighbors. This procedure is equivalent to the Wigner-Seitz
construction known from solid-state physics. In three dimensions the elastic
elements are defined to be the lines of intersection of two neighboring cell
walls, while in two dimensions (see Fig.\ref{fig:vorNet}) they are represented
by the cell walls themselves. We will call these elastic building blocks of the
network \emph{polymer segments} and associate to them the material properties,
respectively the stiffness matrix $\mathbf{K}$, introduced in the preceding
section. By its definition, a segment spans the distance between two vertices
and is therefore ``end-linked'' to the rest of the network.

Depending on the spatial distribution of Voronoi points there will also be a
distribution $P(l_s)$ of segment lengths $l_s$. Only in regular structures, for
example the (anisotropic) two-dimensional honeycomb structure, this distribution
will degenerate into one (or several) delta-function peaks. 

The first moment of this distribution, the average segment length $\lsbar$, is
naturally the most important quantity to describe the geometrical aspects of a
cellular structure. In $d=2,3$ dimensions this ``mesh-size'' may be
reparametrized in terms of the density $\rho$ as
\begin{equation}\label{eq:meshsize}
  \lsbar \propto \rho^{-1/(d-1)}\,,
\end{equation}
where we defined $\rho$ as the total polymer length per system size. While there
are practical reasons to use $\rho$ as a measure for the density in the
simulations, in experimental work it is sometimes easier to control the monomer
concentration $c$. This can be found as $rc\propto\rho$, where the cross-section
radius $r$ is assumed proportional to the monomer size.

\subsection{Mechanical Behavior: Beams}\label{sec:mechanical-behaviour}

In the engineering literature the cellular structures defined above are well
known as foams and are ubiquitous in nature and many areas of technology.
Examples range from liquid foams and froths well known from drinks or household
detergents, to plastic and metallic foams used for insulation or shock
absorption~\cite{wea01,gib99}. It is well known that naturally occurring foams
have to obey Plateau's laws to reach an equilibrium state. We do not require
these laws to hold in the following, since we are interested in the dependence
of elastic properties on the architectural features in general, and not in the
specific details of the dynamic properties of foams.

For purely mechanical cellular foams, where thermal fluctuations are neglected
altogether, the only material length scale is the radius $r$ of the
cross-section. By identifying $\kappa/\lsbar$ as an energy scale, we can use
dimensional analysis to write the shear modulus $G$ as
\begin{equation}\label{eq:scalingGFoam}
  G = \frac{\kappa}{\lsbar^{d+1}} g(r/\lsbar)\,,
\end{equation}
where the occurrence of the spatial dimension $d$ highlights the fact that the
modulus has units of an energy density. In writing this, we have not made
explicit the dependence on the higher moments of the probability distribution
$P$. As will become clear below, these can be used to characterize the
randomness of the structure and will be considered separately. If one defines
force-constants at the scale of the average mesh-size
\begin{equation}\label{eq:barKmech}
  \bar{k}_\perp \simeq \kappa/\lsbar^3\,, \qquad \bar{k}_s \simeq
  \kappa/\lsbar r^2\,,
\end{equation}
the scaling variable can alternatively be written as $r/\lsbar \simeq
\sqrt{\bar{k}_\perp/\bar{k}_s}$, and therefore characterizes the relative
stiffness of the bending to the stretching mode.

\subsection{Regular Structures and Affine Models}\label{sec:regular-foams}

Restricting our attention for the moment to regular structures, macroscopic
elasticity can already be understood by considering the response of a single
cell~\cite{gib99,kra94,zhu97}. In these systems it seems reasonable that local
stresses acting on an individual cell are the same as those applied on the
macroscopic scale. In other words, the local deformation $\delta$ of a cell with
linear dimension $\lsbar$ follows the macroscopic strain $\gamma$ in an \emph{
  affine way} such that it scales as $\delta \propto \gamma \lsbar$. With this
assumption the scaling function can be calculated~\cite{kra94} and one
generically finds for the modulus
\begin{equation}\label{eq:GFoam}
  G_{\rm aff}^{-1} = \lsbar^{d-2} (a\bar{k}_\perp^{-1}+b\bar{k}_s^{-1})\,,
\end{equation}
where the details of the particular structure may enter the numbers $a$ and $b$
in an involved way. The important conclusion to be drawn is that the deformation
modes act as if they were springs connected
in series. For slender beams with $r\ll \lsbar$ the bending mode is softer than
the stretching mode and therefore dominates the modulus -- mechanical foams are
{\it bending dominated}.

While we argue here that the modulus in Eq.(\ref{eq:GFoam}) represents the
generic case, there may be special cases were the prefactors $a$ or $b$ are
suppressed by the specific choice of the unit cell. The triangulated network is
one example where $a=0$ and the bending mode is suppressed. Below we will
encounter another example when studying the square lattice. For these systems
the special geometry of the unit cell, or more generally, the local architecture
has to be taken into account. This is indeed the main focus of this article. On
the other hand, by assuming affine displacements no cooperativity between the
elastic responses of neighboring cells is possible.  The macroscopic modulus $G$
directly reflects the elastic properties of the single cell. The local geometry
is being hidden in the prefactors $a$ and $b$, while the effect of the assembled
structure may simply be predicted by counting the numbers of cells.

\subsection{Cell Polydispersity}\label{sec:cell-polydispersity}

\begin{figure}[tb!]
\begin{center}
  \includegraphics[width=0.9\columnwidth]{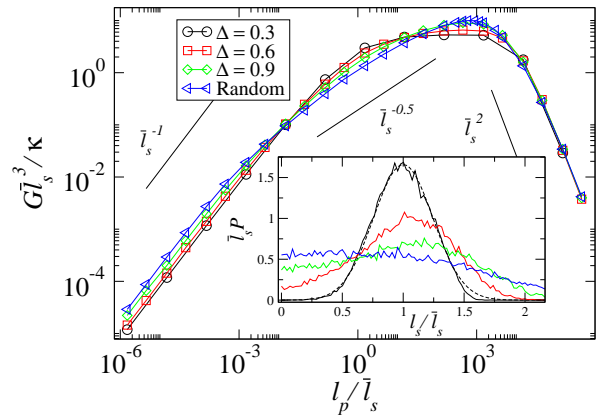}
  \end{center}
  \caption{\label{fig:vorHexMod} Shear modulus $G\lsbar^3/\kappa$ as a function
    of $l_p/\lsbar$ for a 2d honeycomb foam structure with varying degree of
    randomness $\Delta$. The blue curve (``Random'') corresponds to a
    ``maximally'' random foam generated from a Poisson point process. Inset:
    Distribution $P$ of segment lengths for the same systems. At low levels of
    randomness ($\Delta=0.3$) it can be approximated by a Gaussian probability
    distribution (dashed line), while it shows significant broadening upon
    increasing the randomness to $\Delta=0.6,\,0.9$. The peak disappears
    completely in the case of the maximally random Poisson foam. }
\end{figure}

We have tested the validity of the affine model in a simple two-dimensional
cellular structure with varying degree of randomness.  We have taken the seeds
for a Voronoi construction of a regular, honeycomb lattice structure and
randomly displaced them with a uniform probability distribution of width
$\Delta\cdot\lsbar$. The influence of randomness on the elastic properties of
mechanical (non-fluctuating) foams has been studied extensively by various
authors~\cite{sil95,zhu00,faz02}. Here, we also include effects from thermal
fluctuations such that the response of a polymer segment is characterized by
three deformation modes with stiffnesses $k_s,k_\parallel$ and $k_\perp$,
respectively.  The affine prediction for the modulus of this system ($d=2$) can
be inferred from Eq.(\ref{eq:GFoam}). By defining
\begin{equation}\label{eq:barKpar}
  \bar{k}_\parallel\simeq \kappa l_p/\lsbar^4\,,
\end{equation}
and substituting $\bar k_s^{-1} \to \bar k_s^{-1}+\bar k_\parallel^{-1}$ one
finds for the modulus
\begin{equation}\label{eq:GthermFoam}
  G_{\rm aff}^{-1}  = \bar
  k_\perp^{-1} h(l_p/\lsbar) = \frac{\lsbar^3}{\kappa}\left[  a +
    b\left(\frac{R\lsbar}{l_p}\right)^{-2} + c\frac{\lsbar}{l_p}  \right]\,,
\end{equation}
where we have inserted Eqs.(\ref{eq:barKmech}) and (\ref{eq:barKpar}) and used
the relation $R = l_p/r$. This has to be compared with the actual results of our
numerical analysis in Fig.\ref{fig:vorHexMod}. The normalized shear modulus
$G\lsbar^3/\kappa$ is shown as a function of persistence length $l_p/\lsbar$
expressed in units of the average segment length. The curves correspond to
varying degrees of randomness $\Delta$.

We find that regular networks (black curve, circles) characterized by a single
mesh-size $\lsbar$ indeed display the functional form expressed through
Eq.(\ref{eq:GthermFoam}). For mesh-sizes much larger than the persistence length
$\lsbar\gg l_p$ the network deforms by pulling out thermal undulations and
$G\propto \bar k_\perp \propto \lsbar^{-4}$ (left part of
Fig.\ref{fig:vorHexMod}). Decreasing the mesh-size beyond $\lsbar\approx r$
stretching of the polymer backbone dominates the modulus $G\propto \bar
k_s\propto \lsbar^{-1}$ (right part of Fig.\ref{fig:vorHexMod}). The physically
relevant situation for studying stiff polymers, however, corresponds to the
intermediate regime, where the persistence length is much larger than the
mesh-size, which is still much larger than the polymer radius $l_p \gg \lsbar
\gg r$. Typical actin networks with $l_p = 17{\rm \mu m}$ and $r=5{\rm nm}$ may
have mesh-sizes in the sub-micron range $\lsbar\approx 100{\rm nm}$. In this
regime, most of the energy is stored in the bending modes leading to $G\propto
\bar k_\perp \propto \lsbar^{-3}$ corresponding to the plateau region visible in
Fig.\ref{fig:vorHexMod}.

Using the values $a=0.2$, $b=0.35$ and $c=0.14$ we managed to fit the scaling
function of Eq.(\ref{eq:GthermFoam}) to the numerical data (in fact, this is the
dashed line in Fig.\ref{fig:forceconstants}). Increasing the level of randomness
the presence of the additional variable $\Delta$ spoils the scaling property and
a fit is no longer possible. The power law regimes gradually shrink and the
cross-over regions increase in size. While the mechanical stretching regime is
hardly affected by the randomness at all, this effect is most pronounced in the
cross-over from the bending to the thermal stretching dominated regime.  The
physically most relevant intermediate plateau regime disappears completely and
shows strong amplitude modulations.

We have also generated foams by Voronoi tessellation of a fully random
distribution of points, corresponding to a Poisson process (blue curve, left
triangles). For these ``maximally random foams'' one could rather use an
expression $G \propto \lsbar^{-7/2}$ to characterize the modulus at these
intermediate parameter values. At this point this is only an empirical
observation. Later, in the context of the fibrous architecture, we will see how
this exponent can be derived from a scaling argument that properly takes into
account the randomness in the system.

One may infer from the inset of Fig.\ref{fig:vorHexMod} that deviations from the
scaling form presented in Eq.(\ref{eq:GthermFoam}) are indeed intimately
connected to a broadening of the segment length distribution $P(l_s)$. In the
regular structure the distribution can very well be described by a Gaussian
centered around the average mesh-size $\lsbar$ (dashed line in
Fig.\ref{fig:vorHexMod}). Random foams, on the contrary, display significantly
broader distributions and even have non-negligible weight on very small
segments.

We will see below that the different effect of randomness in the thermal and the
mechanical stretching regimes can be traced back to the unusually strong length
dependence of the entropic stretching stiffness $k_\parallel\propto l_s^{-4}$ as
compared to $k_s\propto l_s^{-1}$. We will find that this leads to the breakdown
of the affine model whenever there is a sufficiently broad distribution of
segment lengths. Thermal networks are thus inherently more sensitive to
elements of randomness than purely mechanical systems.

\begin{figure}
\begin{center}  
    \includegraphics[width=0.9\columnwidth]{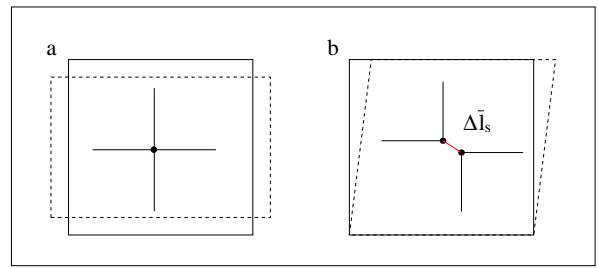}
\end{center}
\caption{ (a) Pure shear deformation of the square lattice and (b) simple shear.
  Illustration of the bifurcation leading from a four-fold connected vertex to a
  three-fold connected one by introducing small amounts of randomness $\Delta\lsbar$.}
  \label{fig:squareShear}
\end{figure}

\begin{figure}
\begin{center}  
  \includegraphics[width=0.9\columnwidth]{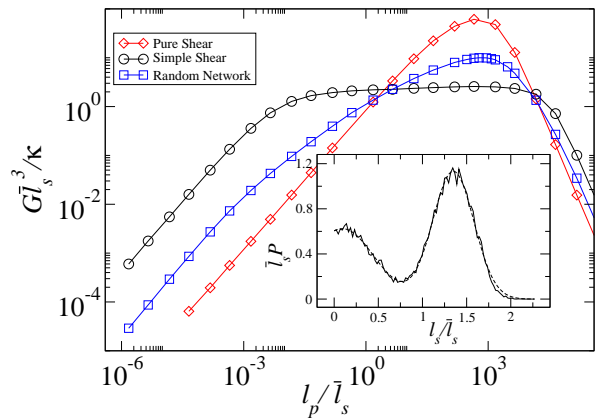}
\end{center}
\caption{The two different shear moduli for the slightly randomized square
  lattice ($\Delta = 0.3$) as shown in Fig.\ref{fig:vorNet}. Also shown is the
  modulus of the highly random Poisson foam. Inset: Distribution
  of segment lengths for the same network. The dashed line is a fit to a sum of
  two Gaussians centered around $l_1/\lsbar=0.092$ and $l_2/\lsbar=1.35$. }
  \label{fig:vorSquareMod}
\end{figure}

It is instructive to consider yet another lattice structure as a basis for our
foam model (see Fig.~\ref{fig:vorNet}). By placing the Voronoi points on a
slightly randomized square lattice one can generate a foam with a bimodal
segment length distribution having a second peak at some small length $l_1$ (see
inset Fig.\ref{fig:vorSquareMod}). To understand this, one has to realize that a
generic foam structure generated by Voronoi tessellation has only three-fold
connected vertices, while they are four-fold connected in the square network. A
small amount of randomness therefore induces a bifurcation of a four-fold vertex
into a short segment with three-fold connected vertices at its ends (see
Fig.\ref{fig:squareShear}).
Unlike the honeycomb foam, the resulting structure is elastically anisotropic
and has $3$ distinct moduli \cite{boa02}. In addition to the bulk modulus there
are two shear moduli corresponding to simple and pure shear deformations. These
two modes are schematized in Fig.\ref{fig:squareShear}, while
the corresponding moduli (together with the isotropic shear modulus of the
Poisson foam) are shown in Fig.\ref{fig:vorSquareMod}.
Pure shear leads to deformations along the main axis of symmetry of the unit
squares and thus to stretching of the elements. The bending regime is therefore
strongly suppressed. On the other hand, simple shear deforms the squares along
their diagonals and thus favors the bending mode. Only when the stretching
energy stored in the small segments $w_\parallel = k_\parallel(l_1)\delta_{\rm
  aff}(l_1)^2 \propto l_p/l_1^2$ equals the bending energy in the average
segment $w_\perp = k_\perp(\lsbar)\delta_{\rm aff}(\lsbar)^2\propto \lsbar^{-1}$
does the system cross-over to a stretching dominated network. Noting (from the
inset of Fig.\ref{fig:vorSquareMod}) that $l_1\approx \lsbar/10$ we find that
this happens when $l_p\approx 10^{-2}\lsbar$ in accord with
Fig.\ref{fig:vorSquareMod}.  It is interesting to see that the network looses
its anisotropy at the two points $\lsbar=l_p$ and $\lsbar=r$, where the modulus
takes the same value as that of the Poisson foam. This follows from the fact
that the stiffness of the average polymer segments is isotropic at these
parameter values and either $\bar k_\parallel \approx \bar k_\perp$ or $\bar k_s
\approx \bar k_\perp$.  Comparing absolute values we find that the shear modulus
in the thermal regime, strongly influenced by the presence of the small
segments, can vary orders of magnitude while the mechanical stretching regime is
hardly affected at all.

To conclude this section we emphasize once again that polydispersity in the
segment lengths can have strong effects on the macroscopic elasticity of a
cellular polymer network. It can lead to modifications of the scaling
properties, as we have found in the most random foams, as well as to
quantitative changes of the modulus by several orders of magnitude as in the
anisotropic square structure. As a consequence, experiments which are limited to
restricted parameter windows would most likely measure effective exponents that
lie in between the extremal values given by pure stretching and bending. One,
therefore, has to be cautious interpreting experimental data within the context
of the foam-model, without the knowledge of the polydispersity of the structure.

\section{Fibrous Architecture}\label{sec:fibrous-architecture}

Looking at pictures of cross-linked actin networks reconstituted in
vitro~\cite{tha06,shi04} one might wonder whether a description in terms of a
cellular architecture is actually relevant for these systems at all. Besides
having a strong polydispersity in cell sizes, real polymer networks seem to have
a hierarchical architecture that allows for smaller cells to be generated within
larger cells within even larger cells. On the contrary, foams only have one of
these hierarchies (see Fig.\ref{fig:vorNet}). What is more, cellular structures
do not account for the effects of the polymer length $l_f$, which constitutes an
additional mesoscopic scale in the problem.

In the following we want to quantify the effects of the polydispersity in
connection with the length scale $l_f$ by studying the elastic properties of a
generic two-dimensional fibrous structure which is defined as follows.  $N$
anisotropic elastic elements, geometrically represented by straight lines of
length $l_f$, are placed on a plane of area $A=L^2$ such that both position and
orientation of the elements are uniformly random distributed.  {\boldface This randomness
entails a distribution of angles $\theta\,{\rm\epsilon}\,[0,\pi]$ between two
intersecting filaments
\begin{equation}\label{eq:angleDist}
  P(\theta) = \frac{\sin(\theta)}{2}\,,
\end{equation}
which has a maximum for filaments at right angles.} At any intersection a
permanent cross-link with zero extensibility is generated.  This constrains the
relative translational motion of the two filaments. For the rotational degree of
freedom one may introduce an energy contribution $W_{\rm rot} =
m(\phi-\phi_0)^2$ for the change of relative cross-link angles $\phi$ from their
initial values $\phi_0$.  We restrict ourselves to the study of the two limiting
cases, where the potential is either soft ($m\to0$) and therefore allows for
free relative rotations of the filaments (free hinges), or infinitely stiff
($m\to\infty$) and inhibits any change of the angles at the cross-links (fixed
angles).

The remaining elastic building blocks of the network, the \emph{polymer
  segments}, span the distance between two neighboring cross-links on the same
polymer. Their length can be shown to follow an exponential distribution
\cite{kal60}
\begin{equation}\label{eq:segDist}
  P(l_s)=\lsbar^{-1} e^{-l_s/\lsbar}\,.
\end{equation}
The mean value $\lsbar$ is given in terms of the density $\rho=Nl/A$ as 
\begin{equation}\label{eq:rholsbar}
\lsbar=\pi / 2\rho\,,
\end{equation}
which is a realization of Eq.(\ref{eq:meshsize}). On average there are, thus,
$x=l_f/\lsbar \approx l_f\rho$ segments per polymer. The simplicity of this
network, which has only one structural parameter $\rho$, makes it an ideal
candidate to obtain physical insight into the relation between architecture and
elastic properties of the constituents. This model has frequently been used to
study the elastic and brittle properties of athermal paper
sheets~\cite{ast00,ast94,lat01,hey96}. In the context of biological networks of
stiff polymers it has been introduced in \cite{frey98} and recently studied by
various authors \cite{wil03,hea03a,hea03c}. In all this work, however, the
elastic properties of the polymers are modeled by the classical theory of
Euler-Bernoulli beams. Here, we concentrate on the effects of thermal
fluctuations, a brief account of which we have published recently~\cite{heu06a}.

\subsection{Simulation Results}\label{sec:simulation-results}

\begin{figure}
  \begin{center}
    \includegraphics[width=0.9\columnwidth]{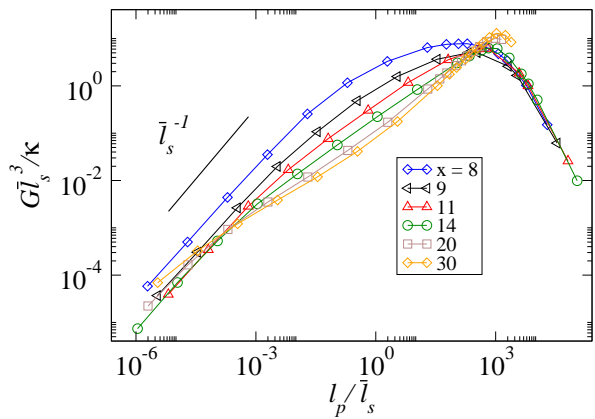}
  \end{center}
  \caption{Scaling function $g$ as a function of $y=l_p/\lsbar$ for various
    $x=l_f/\lsbar\le30$ (from networks with freely hinged cross-links); For
    rather small values of $x=8,9$ the curves resemble the results from the
    cellular networks. At intermediate values $10^{-3}\le y \le 10^2$ the
    modulus shows strong modulation and develops a dip with increasing $x$.}
  \label{fig:modulusScaledSmall}
\end{figure}

\begin{figure}
  \begin{center}
    \includegraphics[width=0.9\columnwidth]{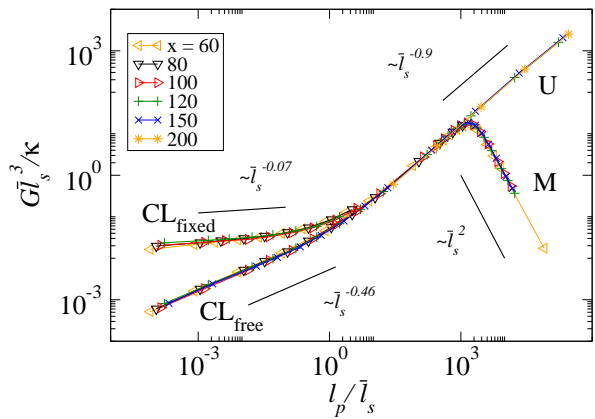}
  \end{center}
  \caption{Scaling function $g$ as a function of $y=l_p/\lsbar$ for various
    $x=l_f/\lsbar\ge60$; In contrast to Fig.\ref{fig:modulusScaledSmall} no
    dependence on $x$ is observed any more and new scaling regimes emerge. The
    two branches in the cross-link dominated regime ($y\ll1$) correspond to
    freely hinged ($\rm CL_{free}$) or fixed ($\rm CL_{fixed}$) cross-link
    angles, respectively. For $y\gg1$ one encounters a universal thermal regime
    (U) independent of the cross-link properties as well as a mechanical regime
    (M).}
  \label{fig:modulusScaled}
\end{figure}

In Figs.~\ref{fig:modulusScaledSmall} and \ref{fig:modulusScaled} the results of
our simulations are shown for fibrous networks with a varying number $x$ of
cross-links per polymer. The axes are the same as in previous plots. The
normalized shear modulus $G\lsbar^3/\kappa$ is shown as a function of
persistence length $y=l_p/\lsbar$ expressed in units of the average segment
length.
Short fibers with few cross-links, corresponding to low densities are depicted
in Fig.~\ref{fig:modulusScaledSmall}, long fibers or high densities in
Fig.~\ref{fig:modulusScaled}. In both figures we find a regime at large values
of the persistence length $l_p/\lsbar$ (right part of the plot) where the
dimensionless shear modulus decreases as $G\propto l_p^{-2}\propto r^{-2}$. This
corresponds to a purely mechanical stretching regime where $G\propto\bar{k}_s$ consistent
with the mean-field picture of Eq.(\ref{eq:GthermFoam})
\cite{frey98,wil03,hea03a,ast00}.

Our main interest, however, lies in the regime of $l_p/\lsbar \leq 10^3$, where
the persistence length is small enough for thermal fluctuations to become
relevant.  In this regime one may safely neglect the mechanical stretching
stiffness and set $k_s\to\infty$. Then, dimensional analysis for the shear
modulus requires
\begin{eqnarray}\label{eq:scalingGFiber}
  G = \frac{\kappa}{\lsbar^{3}} g(x, y)\,,
\end{eqnarray}
where we have introduced the scaling variables
\begin{eqnarray}\label{eq:abbrev}
  x = l_f/\lsbar\simeq l_f\rho\,, \qquad y = l_p/\lsbar\simeq l_p\rho\,.
\end{eqnarray}
Comparing to Eq.(\ref{eq:scalingGFoam}) there is an additional argument in the
scaling function $g$, the polymer length $x=l_f/\lsbar$. This purely geometrical
variable counts the number of cross-links (or equivalently: segments) per
filament. The second argument may be written in the alternative form $y \simeq
\bar{k}_\parallel/\bar{k}_\perp$. It characterizes the relative stiffness of
stretching and bending mode for a typical segment.

From Fig.~\ref{fig:modulusScaledSmall} one infers that for \emph{low densities}
$g = y f(x)$, implying for the modulus $G=\bar{k}_\parallel f(\rho l_f)$.  This
linear dependence on the ``pre-averaged'' stretching compliance
$\bar{k}_\parallel$ hints at an entropic stretching dominated regime similar to
that found in the cellular structures discussed above. This regime has been
suggested in \cite{hea03c}, where a scaling argument is developed relying on the
affine assumption borrowed from the mechanical stretching regime.  Our analysis
shows that the domain of validity of this linear regime is extremely narrow and
confined to short filaments $x \leq 20$ and persistence lengths $y \ll 1$. What
is more, due to the non-trivial density-dependence expressed through the
function $f(x)$, the modulus does not even display a power-law behavior in the
density. Instead, we find that the modulus shows complex dependence on its
variables and develops a dip in the intermediate parameter region where
$10^{-3}\le y \le 10^2$. This is also the relevant parameter range for networks
of F-actin, where the ratio of persistence length to mesh-size $l_p/\lsbar
\approx 10-100$.

For \emph{medium and high densities} Fig.~\ref{fig:modulusScaled} shows
non-trivial scaling regimes where the scaling function $g$ becomes independent
of $x$ and therefore of the filament length $l_f$. This highly non-trivial
observation has important implications and allows the system to exhibit power
law behavior $g\propto y^z$. We find non-trivial fractional exponents $z=0.46
(0.07)$ and $z=0.9$ for small and large values of $y$, respectively. In the
figure one can distinguish four branches that belong to different realizations
of the network. While branch M (mechanical regime, $G\propto\bar k_s$) has been
discussed already, the remaining three are obtained by setting $k_s\to\infty$.
The two branches found at small values $y\ll 1$ relate to networks where the
cross-link angles are either free to vary ($\rm CL_{free}$, $z=0.46$) or are
perfectly fixed to their initial values ($\rm CL_{fixed}$, $z=0.07$),
respectively.  We term this regime ``cross-link dominated'' since tuning the
cross-link properties may have strong effects on the elastic modulus by driving
the system from one branch towards the other. Both branches merge at $y\approx
1$ where we enter a universal regime (branch U, $z=0.9$) which is completely
independent of the elasticity of the cross-links and which therefore is termed
``filament dominated''.
\begin{table}
  \begin{tabular}{|c|c|c|c|c|}\hline
    &  & $z$ & $\;z_{\rm Theory}\;$ & $\;z_{\rm Foam}\;$ \\ \hline
    $\rm CL_{fixed}\;$ & $\;r\ll l_p\ll \lsbar\;$ & $\;0.07$ & 0 & $1$  \\ \hline
    $\rm CL_{free}\;$ &   $\;r\ll l_p\ll \lsbar\;$ & $\;0.46$ & 1/2 & --  \\ \hline
    $\rm U$ & $\;r\ll\lsbar\ll l_p\;$ & $\;0.9$ & $1$ & 0 \\ \hline
    $\rm M$ & $\;\lsbar \ll r\ll l_p\;$ & $\;(1)$ & $(1)$ & $(1)$ \\\hline
\end{tabular}

\caption{\label{tbl:regimes} {\boldface Compilation of the different elastic regimes of the fibrous
    network. The modulus is given by $G\sim \bar k_\perp^{1-z} \bar k_\parallel^z$
    with the appropriate values for the exponent $z$. For comparison also the
    predictions from the theoretical analysis (see below) as well as the 
    exponents for the foam structure are given. The latter only for fixed
    cross-link angles ($\rm CL_{fixed}$), which is necessary to 
    make the structure stable. The mechanical regime $\rm M$ corresponds to the exponent
    $z=1$, however with $\bar k_\parallel$ substituted by $\bar k_s$.}} 
\end{table}

 In all cases, the modulus can be written as a generalized geometric average
\begin{equation}\label{eq:GPoly} 
  G \propto \bar{k}_\perp^{1-z}\bar{k}_\parallel^z\,,
\end{equation}  
which has to be contrasted with Eq.(\ref{eq:GthermFoam}), where bending and
stretching modes are assumed to superimpose linearly {\boldface(see
  Table~\ref{tbl:regimes} for a direct comparison of the various regimes)}.
There, the system is described either by $z=0$, if bending dominates, or by
$z=1$ if stretching is the main deformation mode.  Values different from the two
limiting cases $z=0,1$ cannot be described by the mean-field approach, hence the
assumption of affine deformations applied on the level of the polymer segments
(or the cell size) necessarily has to fail. This will become especially clear in
the following section, where we review the application of affine theories to
fibrous architectures. We will illustrate its failure and highlight the physical
principles involved. To go beyond we will introduce a model that accounts for
the spatial distribution of cross-links along the backbone of a \emph{typical
  polymer filament} instead of just considering a single \emph{typical polymer
  segment}. This new approach will allow us to understand all the features of
the macroscopic elasticity encountered in Fig.\ref{fig:modulusScaled}.

\subsection{Affine Models in Fibrous Architectures}\label{sec:affine-model}

In some of the earlier approaches to describe the elastic moduli of stiff
polymer networks the assumption of affine deformations has been applied on the
level of the average segment which can be characterized by ``pre-averaged''
response coefficients $\langle k(l_s)\rangle \to \bar{k} = k(\lsbar)$ introduced
in Eqs.(\ref{eq:barKmech}) and (\ref{eq:barKpar}). The characteristic fibrous
structure of stiff polymer networks is not accounted for and effectively
substituted by a highly regular cellular structure. The modulus in the thermal
regime is then obtained simply by replacing in Eq.(\ref{eq:GFoam}) the
mechanical stretching response $\bar{k}_s$ with its thermal counterpart
$\bar{k}_\parallel$. Several variants of this model have been considered in the
literature \cite{mac95,kro96,sat96} that only differ in the specific (ad hoc)
choice of the prefactors $a,b$. The stretching dominated model \cite{mac95}
(setting $a=0$ in Eq.(\ref{eq:GFoam})) with a modulus depending on density as
\begin{equation}\label{eq:GAffSt}
G_\parallel\sim \rho^{(2+d)/(d-1)}\,,
\end{equation}
and its extensions to nonlinear elasticity \cite{sto05}, have widely been used
to fit experimental data for the plateau modulus in cross-linked F-actin
networks \cite{gar04b,shi04,tha06}. Despite this apparent success, it is not clear
a priori why in the parameter regime of interest the mesh-work should deform by
the stretching of bonds when actually bending is by far the softer mode
($\bar{k}_\parallel/\bar{k}_\perp\simeq l_p/\lsbar\gg 1$). In general, such a
regime can only occur if the specific architecture suppresses the soft bending
modes as in the triangulated structure with its highly coordinated vertices. A
second approach seems to repair this deficiency by setting in
Eq.(\ref{eq:GFoam}) $b=0$. The modulus in this theory
\begin{equation}\label{eq:GAffBend}
  G_\perp \sim \rho^{(1+d)/(d-1)}\,,
\end{equation}
only differs by a factor of $\rho^{1/(d-1)}$ from the stretching dominated
modulus of Eq.(\ref{eq:GAffSt}). However, neither theory provides justification
for neglecting the effects of the polydispersity in the fibrous system. In fact,
if one extends the approach to include the distribution of segment lengths, such
theories necessarily have to fail, as we will explain in the next section.

\begin{figure}
\begin{center}
  \includegraphics[width=0.8\columnwidth]{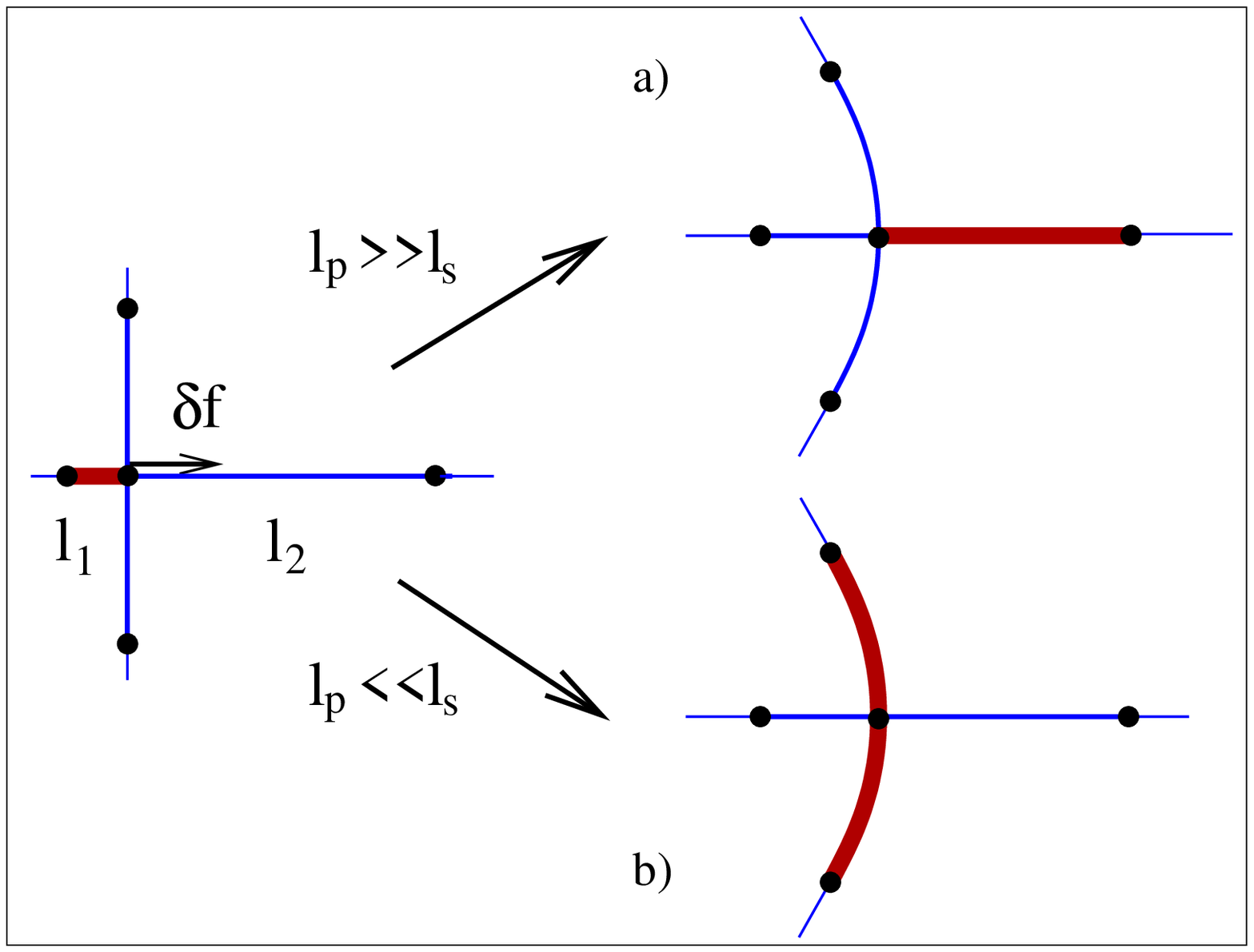}
\end{center}
\caption{Illustration of the effects of non-zero residual forces. The relaxation
  of the small segment $l_1$ from its overly stressed state goes to the cost of
  additional deformations in its neighbors. Depending on the value of the
  persistence length the energy will mainly be stored in a) the stretching mode
  ($l_p \gg \lsbar$) or b) the bending mode ($l_p \ll \lsbar$).}
  \label{fig:mechanism}
\end{figure}

\subsection{Effects of the Segment Length
  Distribution}\label{sec:length-distribution}

To understand the origin of this failure consider an affine deformation field
$\delta_{\rm aff}\propto\gamma l_s$ being applied to a random network of stiff
polymers with a distribution $P(l_s)$ of segment lengths $l_s$. The axial forces
$f_\parallel$ generated by such a deformation field can simply be calculated by
multiplying the deformation with the stretching stiffness of the segment
\begin{equation}\label{eq:fAff}
f_\parallel = k_\parallel \delta_{\rm aff} \simeq \kappa l_p\gamma / l_s^3\,.
\end{equation}
Note that in contrast to the purely mechanical situation, where the axial force
$f_s = k_s\delta_{\rm aff} \simeq \kappa\gamma/r^2$ is independent of length,
$f_\parallel$ strongly increases with shortening the segment length.  This
implies that, in general, two neighboring segments on the same filament produce
a net force $\delta\!f$ at their common node that has to be taken up by the
crossing filament. There, it leads to additional deformations that eventually
destroy the affine order. This mechanism is illustrated in
Fig.\ref{fig:mechanism} where the relaxation of the small segment $l_1$ leads to
bending of its neighbor on the crossing filament (b). Also the segment $l_2$ on
the same filament is affected by the relaxation and experiences an
additional stretching contribution (a). Whether the available energy is stored in
the stretching or the bending mode depends crucially on the value of the
persistence length, as is indicated in the figure.

One may calculate the probability distribution $Q$ for residual forces by
summing over all segment lengths that are consistent with a given force
$\delta\!f$,
\begin{equation}\label{eq:resForce}
  Q(\delta\!f) = \left\langle \delta(|f_\parallel(l_1) -
    f_\parallel(l_2)| - \delta\!f) \right\rangle\,.
\end{equation}
The averaging procedure defined by the angular brackets
\begin{equation}\label{eq:angBrackets}
  \langle A\rangle = \int dl_1\!\int dl_2P(l_1,l_2)\,A(l_1,l_2)\,,
\end{equation} 
involves the two-point probability $P(l_1,l_2)$ of finding neighboring segments
with lengths $l_1$ and $l_2$, respectively. In the special case of the random
network considered here, there are no correlations between neighboring segment
lengths such that the distribution factorizes. The formula can thus be evaluated
by substituting $f_\parallel\propto l_s^{-3}$ taken from Eq.(\ref{eq:fAff}).
This inverse relationship between forces and segment lengths translates the
weight of the probability distribution $P_0=P(l_s\to 0)\neq 0$ at small segment
lengths into polynomial (fat) tails of the corresponding distribution of
residual forces
\begin{equation}
  Q(\delta\!f\to\infty)\propto \delta\!f^{-4/3}P_0\,,
\end{equation}
which has a diverging mean value. The exponent can readily be derived from
evaluating the integral measure $df_\parallel \propto l^{-4}dl \propto
f_\parallel^{4/3}dl$. As a consequence there are always residual forces high
enough to cause additional deformation of the crossing filament. Hence we
conclude that an affine deformation field is unstable and that the system can
easily lower its energy by redistributing the stresses to relieve shorter
segments and remove the tails of the residual force distribution $Q(\delta\!f)$.

Even though we have evaluated Eq.(\ref{eq:resForce}) for the special case of an
exponential segment length distribution Eq.(\ref{eq:segDist}), it is important
to note, that the observed sensitivity is not a special feature of the fibrous
architecture, but applies to any polymer network with a broad distribution of
segment lengths {\boldface independent of the dimensionality of the network}.
Due to the strong length dependence of $k_\parallel(l_s)$ the thermal response
is highly sensitive to even small polydispersity as we have already seen in the
random cellular network of Sect.\ref{sec:cell-polydispersity}. On the contrary,
these effects are completely absent in purely mechanical models and also in
models of flexible polymers, where the distribution $Q(\delta\!f)$ degenerates
into a delta-function peak at the value $\delta\!f=0$, and explains the
robustness of these regimes to randomness.

If we want to include the effects of randomness into a microscopic theory we
cannot naively apply the conventional picture of affine deformations on the
scale of the single segment. This can safely be done only in highly ordered
structures like regular cellular materials. Instead, we have to adopt a
description of the deformations (at least) on the larger scale of the complete
polymer.  In the following we therefore consider a \emph{typical polymer
  filament}, which is composed of a sequence of segments drawn from the
distribution $P(l_s)$. To describe the elastic properties correctly, we will
also have to consider the connections of the polymer to the surrounding network
matrix, in addition to the elastic properties of the segments themselves. We may
now employ this picture to explain the intricate scaling properties of the
polymer network in all the parameter regimes displayed in
Fig.\ref{fig:modulusScaled}.

\subsection{Cross-link dominated Regime}\label{sec:freely-hinged-cross}

\subsubsection{Freely hinged cross-links}

We start with the description of the system in the parameter region $y\ll1$
($l_p \ll \lsbar$), where the properties of the cross-links strongly influence
the system's response. The idea is to impose a virtual affine deformation on
every segment and calculate, as a perturbative correction, the contribution to
the elastic energy resulting from the relaxation out of this reference state.
This procedure will lead to good predictions only when the corrections are small
and the affine deformations are only weakly perturbed. As we will see below,
this is the case in the parameter region $y\ll1$. However, it will also become
clear, that a small perturbation for the deformations is sufficient to generate
completely different scaling properties for the macroscopic modulus. For the
moment we restrict our attention to free relative cross-link rotations (branch
$\rm CL_{fixed}$), since then the affine reference state is particularly simple
and contains stretching contributions only.

As explained above any deviation from the affine reference state, induced by
relaxation of non-zero residual forces, will lead to additional deformations in
the crossing filaments. Since it is more likely that two filaments cross each
other at an angle close to $90^\circ$, the induced non-affine deformations will
mainly be oriented transverse to the contour of the crossing filament and are
therefore of bending character. The value of the exponent $z=0.46$ supports this
assumption and indicates that bending and stretching deformations in this regime
contribute equally to the elastic energy even though the bending mode is very
stiff ($\bar k_\parallel /\bar k_\perp \sim y\ll 1$). Therefore any relaxation
of residual stretching forces, will be punished by high amounts of bending
energy (see Fig.~\ref{fig:mechanism}b).  Only the smallest segments on the
polymer, corresponding to the outermost tails of the residual force
distribution, will have sufficient energy to perturb the deformation field and
relax to a state of lower strain.

In the following, we will assume that segments up to a critical length $l_c$ --
to be determined self-consistently -- fully relax from their affine reference
state to give all their energy to the neighboring segment on the crossing
filament. The total energy of the polymer can then be calculated from segments
with $l_s>l_c$ only. There are two contributions. First, a stretching energy
\begin{equation}\label{eq:strEnAff}
  w_s(l_s) \simeq k_\parallel\delta_{\rm aff}^2 \simeq 
\kappa\gamma^2\frac{l_p}{l_s^2}\,,
\end{equation}
from the imposed affine strain field $\delta_{\rm aff} \propto \gamma l_s$.
Second, a bending energy
that is due to the relaxation of a neighboring segment on the crossing filament
out of its affine reference state. This process requires that the segment of length
$\hat l_s$  moves the distance $\hat\delta_{\rm aff} = \gamma \hat l_s$, which
corresponds to its own affine deformation. The resulting bending energy
\begin{equation}\label{eq:bendEn}
  w_b(l_s) \simeq  k_\perp\hat\delta_{\rm aff}^2 \simeq
  \kappa\gamma^2\frac{\hat{l}^2_s}{l_s^3} \, ,
\end{equation}
now depends on the length $l_s$ of the segment under consideration as well as on
the length $\hat l_s$ of the neighboring (now relaxed) segment. As we have
assumed above, the second contribution $w_b$ only arises if the length $\hat
l_s$ is shorter than the critical length $l_c$. The total deformation energy
along the polymer is then obtained by adding both contributions and integrating
over all segments $l_s>l_c$ along the filament as well as averaging over
neighbors with $\hat{l}_s<l_c$ ,
\begin{equation}\label{eq:totEn}
  W_{\rm pol} \simeq (l_f\rho)\kappa\gamma^2\int_{l_c}^\infty dl_sP(l_s)\left( \frac{l_p}{l_s^2} +
    l_s^{-3}\int_0^{l_c}d\hat{l}_sP(\hat{l}_s)\hat{l}^2_s \right)\,,
\end{equation}
where the prefactor $l_f\rho$ just counts the number of segments per polymer.
For simplicity, we have not considered any dependence of the deformations on the
orientation relative to the macroscopic strain field. In essence, this would
only introduce some additional numerical prefactors that are irrelevant for the
scaling picture developed here. The integrations are reparametrized by
introducing the non-dimensional variable $\lambda=\rho l_s$ such that we arrive
at the expression for the average polymer energy
\begin{equation}\label{eq:totEn2}
  W_{\rm pol}\simeq  \kappa\gamma^2l_f\rho^2(\rho l_p/\lambda_c+\lambda_c)\,,
\end{equation}
where numerical constants have been dropped and $\lambda_c := \rho l_c \ll 1$ in
the parameter range of interest.  Minimizing with respect to $\lambda_c$
determines a new non-affinity length
\begin{equation}\label{eq:naLength}
  l_c^{\rm min} = \lambda_c^{\rm min}\lsbar \simeq \sqrt{l_p\lsbar}\,,
\end{equation}
that sets the maximal scale up to which the destruction of affine deformations
lead to a lowering of the elastic energy. Inserting this length into
Eq.(\ref{eq:totEn2}) and multiplying by the number-density of filaments
$\rho/l_f$ one arrives at an expression for the modulus $G\simeq W_{\rm
  pol}^{\rm min}\cdot\rho/l_f\gamma^2 \simeq \kappa\rho^{7/2}l_p^{1/2}$.
Rewriting the result as
\begin{equation}\label{eq:modGeoAvg}
  G \simeq \sqrt{\bar{k}_\perp\bar{k}_\parallel} \;\propto\; \lsbar^{-7/2}\,,
\end{equation} 
we immediately see that our theory reproduces the empirical result of
Eq.(\ref{eq:GPoly}) with an exponent $z=1/2$, which compares
well with the measured value of $z=0.46$. 

\begin{figure}
\begin{center}
  \includegraphics[width=0.9\columnwidth]{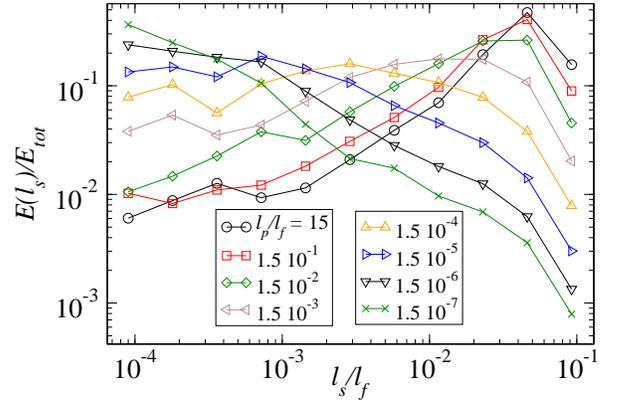}
\end{center}
\caption{Fraction of energy stored in the various segment lengths;
  the curves correspond to different persistence lengths at a density
  of $l_f/\lsbar=80$, equivalent to $\lsbar/l\approx 2\cdot10^{-2}$.}
  \label{fig:enRatio}
\end{figure}

The non-trivial behavior of $G$ observed in Fig.~\ref{fig:modulusScaled} can
thus be explained by a non-affinity length scale $l_c^{\rm min} \simeq
\sqrt{\lsbar l_p}$ below which the affinity of the deformation field breaks
down.  Recapitulating the results from the cellular networks in
Fig.\ref{fig:vorHexMod}, we observe that the same intermediate scaling behavior
of $G\propto \lsbar^{-7/2}$ is found in both architectures. We have thus
established the microscopic origin of the scaling law. It derives from a
continuous unloading of smaller segments driven by an interplay between segment
length distribution and elastic properties of the single polymer. This mechanism
is illustrated in Fig.~\ref{fig:enRatio}, where a histogram for the fraction of
energy stored in segments of various lengths is shown.  For very small
persistence length, a significant fraction of the energy is stored in the
shortest segments. Affine deformations can be seen as a good approximation.
Increasing the persistence length, the short segments one after the other loose
their energies in favor of additional excitations in longer segments.  This is
fully consistent with the assumption of a growing non-affinity scale $l_c^{\rm
  min}$ below which no energy is stored.

It is important to realize that our derivation of the exponent does not make use
of the precise form of the segment length distribution $P(l_s)$. In fact, there
is no need to perform the integrations explicitly and only the limiting
behavior of $P(l_s\to0)$ enters. Thus, the conclusions are valid for a general
class of functions that may even be slowly vanishing at zero segment length.

\begin{figure}
\begin{center}
  \includegraphics[width=0.9\columnwidth]{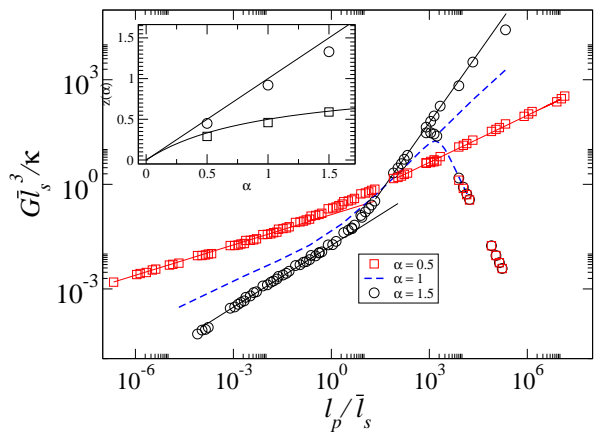}
\end{center}
\caption{Scaling function $g$ as a function of $y=l_p/\lsbar$; the symbols
  correspond to values of $\alpha=0.5,1.5$. In addition the scaling function for
  $\alpha=1$ taken from Fig.\ref{fig:modulusScaled} is shown (dashed line).
  Inset: Exponents $z$ determined from the slopes of the branches $\rm
  CL_{free}$ (squares) and U (circles), respectively. The solid lines correspond
  to the curves $z=\alpha$ and $z=\alpha/(1+\alpha)$ as derived in the main
  text.  }
  \label{fig:genExponents}
\end{figure}

We have also conducted simulations that assume a more general form for the
stretching stiffness
\begin{equation}\label{eq:gen_k_entropic}
k_\parallel(\alpha) = 6\kappa\frac{l_p^\alpha}{l_s^{3+\alpha}}\,,
\end{equation} 
which reduces to the original definition for $\alpha=1$. Since the relative
stiffness of the deformation modes is now $k_\parallel/k_\perp\propto
(l_p/l_s)^\alpha$ we can think of the phenomenological exponent $\alpha$ to tune
the anisotropy of the individual segment. It allows us to extend our discussion
to the broad class of systems for which $k_\parallel$ is a monomial (with units
energy per area) involving one additional material length $l_p$. Repeating the
scaling theory for general values of $\alpha$ gives
$z(\alpha)=\alpha/(1+\alpha)$ which is verified by the results of the
simulations presented in Fig.\ref{fig:genExponents}. It provides further
evidence for the validity of our scaling picture.

\subsubsection{Fixed Cross-link Angles}

If we want to apply the same reasoning to the network with the fixed cross-link
angles, we face the problem that even a perfectly affine displacement of all the
cross-links induces some amount of bending of the segments, in addition to the
usual contribution from the stretching deformations. While an affine strain
$\gamma$ would change all angles by an amount $\Delta\phi\propto\gamma$, due to
the infinite rotational stiffness in the cross-links this cannot actually occur.
The segments therefore have to experience an extra bending contribution induced
by cross-link rotations $(-\Delta\phi)$ that restore the angles to their
original values. In the parameter regime $y\ll1$, where bending is the stiffer
mode, we therefore expect strong contributions to the energy from the bending
mode already in the affine reference state. Allowing for the relaxation of the
smallest segments from their stretched state to even stronger reduce the amount
of stretching we might find an exponent as low as $z=0.07$, signalling nearly
exclusive contributions from the bending mode, not too surprising. In fact, we
will argue below that neglecting the stretching energies, i.e. assuming an
exponent of $z=0$, represents a reasonable approximation to the elasticity in
this regime.

\subsection{Universal Regime}\label{sec:universal-regime}

By increasing $y$ from its small value we soften the bending mode and therefore
reduce the influence of the fixed cross-link angles on the elastic energy. At
the same time the non-affinity scale $l_c^{\rm min}\propto y^{1/2}$ increases,
indicating ever stronger deviations from the affine reference state. When,
eventually, $l_c^{\rm min}\approx l_p\approx\lsbar$ ($\lambda_c^{\rm min}
\approx y \approx 1$) the affine strain field does not serve as a reference
configuration any more, since it is strongly perturbed by a majority of segments
with $l_s<l_c^{\rm min}$. At this stage, the two branches, present in the
cross-link dominated regime, merge and one enters a universal filament-dominated
regime. There, the specific properties of the cross-links do not influence the
macroscopic elasticity notably.

While the scaling argument presented for the hinged network ceases to be
applicable, the remaining residual forces $\delta\!f$ continue to lead to a
redistribution of stresses from shorter to longer segments, albeit at higher
scales. As we have shown in \cite{heu06a}, eventually about $90\%$ of the energy
is stored in the longest $30\%$ of the segments only. The new feature as
compared to the regime $\rm CL_{free}$ is that unloading of a segment from its
stretched configuration will also lead to stretching of its immediate neighbor
on the {\em same} filament (see Fig.\ref{fig:mechanism}a).  This way, the
available energy for bending of the {\em crossing} filament, which was the
primary contribution in Eq.~(\ref{eq:totEn}), is reduced. In the limit $y\gg1$
we can neglect these contributions and calculate the energy from the polymers'
stretching stiffness only. The physical picture is that of a serial connection
of infinitely many segments along the backbone of a ``typical'' polymer. The
stiffness of this polymer, and therefore the modulus, may be obtained from the
stretching spring constants of the individual segments $k_\parallel(l_s)$ as
\begin{equation}\label{eq:infserial}
  G^{-1} = \int dl_s P(l_s) k_\parallel(l_s)^{-1}\propto \bar{k}_\parallel^{-1}\,,
\end{equation} 
corresponding to the exponent $z=1$.  For the more general response coefficient
of Eq.(\ref{eq:gen_k_entropic}) this argument predicts $z=\alpha$, a result
which is closely confirmed by the results of the simulation as can be seen from
the inset of Fig.\ref{fig:genExponents}.  Note, that the shear modulus in this
asymptotic region takes the same form as postulated by the affine theory in
Eq.(\ref{eq:GthermFoam}). However, using Eq.(\ref{eq:infserial}) one can resolve
the effects according to segment length to find that the contribution to the
total energy from segments with length $l_s$ grows as $W(l_s)\propto l_s^4$.
This strong increase is in accord with the assumption of a large non-affinity
scale, below which no energy is stored, and in striking contrast to the affine
theory that would yield $W_{\rm aff}(l_s) \propto k_\parallel(l_s)\delta_{\rm
  aff}^2\propto l_s^{-2}$.

\section{Conclusion}\label{sec:conclusion}

We have studied the macroscopic elastic properties of networks of semi-flexible
polymers. We provide exhaustive numerical studies supplemented by scaling
arguments that elucidate the subtle interplay between the architecture of the
network and the elastic properties of its building blocks.

The main conclusion to be drawn is that, irrespective of the specific
architecture chosen, thermally fluctuating stiff polymer networks are inherently
more sensitive to polydispersity and randomness than their purely mechanical
counterparts. This is due to their strongly length-dependent entropic stretching
response $k_{\parallel} (l) \propto \kappa^2/l^4$ which has to be contrasted
with the mechanical stretching stiffness $k_s(l)\propto\kappa/l$. 

Although simulations have only been conducted in two-dimensional networks, the
identified mechanism by which the structural randomness influences the elastic
properties is expected to be of universal character and hold independent of
dimensionality.  As we have shown, the actual consequences of this
susceptibility (e.g. scaling behaviour of elastic moduli) may vary from system
to system and certainly also with the dimension. A precise knowledge of the
network architecture is therefore indispensable for the interpretation of
experimental data. For this it will be most important to develop new techniques
that allow the characterization of the microstructure and monitor its changes
upon deformation. As exemplified by the discussion in the universal regime
Sec.~\ref{sec:universal-regime}, where the (non-affine) elastic modulus turns
out to be similar to that in an affine theory, we have shown that macroscopic
measurements alone do not suffice to extract the network mechanics also on the
microscopic scale.

We have described how the polymer length $l_f$ can be used to drive the system
from a simple cellular structure with filaments as short as the mesh-size
$l_f\approx \lsbar$, to a fully scale-invariant fibrous structure characterized
by infinitely long filaments $l_f\to\infty$. Especially the latter limit allows
for intricate scaling behavior that impressively demonstrates the qualitative
difference between thermally fluctuating and purely mechanical elastic networks.

The elasticity of a simple cellular structure may be described by a serial
connection of their elementary deformation modes bending and stretching,
respectively. This leads to the modulus of Eq.(\ref{eq:GFoam})
\begin{equation}
  G^{-1} = a\bar k_\perp^{-1} +b\bar k_\parallel^{-1}\,.
\end{equation}
In this picture, deformations can be drawn from either mode and it will be the
softer one that dominates the modulus. In fibrous networks with fixed cross-link
angles we have shown that the modes rather act as if they were springs connected
in parallel. The modulus can then be approximated by
\begin{equation}\label{eq:Gparallel}
  G = a\bar k_\perp +b\bar k_\parallel\,,
\end{equation}
where the prefactors $a,b$ depend weakly on the scaling variable
$y\sim\bar k_\parallel/\bar k_\perp\sim l_p/\lsbar$. The network elasticity is
therefore always dominated by the stiffer mode, qualitatively similar to a
triangulated network, where the specific geometry of the unit cell always
imposes stretching deformations on the system, no matter how soft the bending
mode actually is. The fibrous architecture apparently also suppresses the
transition into regimes where the softer mode is dominant. This conclusion is
consistent with recent simulations on the purely mechanical fiber
model~\cite{wil03,hea03a}, where a transition into a regime dominated by soft
bending modes ($y\gg1$) could only be observed at finite values for the filament
length $l_f$. Increasing the length to asymptotic values $l_f\to\infty$, as we
have done here, such a ``bending-soft'' regime is strongly suppressed and
eventually cannot occur any more. Instead, the elasticity is governed by the
much stiffer (mechanical) stretching mode. A detailed theoretical explanation of
how this suppression is generated in mechanical fiber networks will appear
elsewhere~\cite{heu06b}, however, it is clear that the mechanism that leads to
bending in cellular structures cannot work in fibrous networks. The fact, that
any segment is part of the larger structure of the polymer fiber leads to strong
geometric correlations and imposes very strict conditions on possible segmental
deformations.

Allowing the filaments to freely rotate at the cross-links, a situation which
may be relevant for F-actin networks cross-linked for example with
$\alpha$-actinin, we also find an asymptotic scaling regime where stretching and
bending modes contribute equally to the elastic energy, Eq.(\ref{eq:GPoly}),
\begin{equation}\label{} 
  G \propto \bar{k}_\perp^{1-z}\bar{k}_\parallel^z\,.
\end{equation}  
By quantifying the degree of co-operation between neighboring elements in the
network we were able to identify a non-affinity length-scale $l_c$ below which
the state of affine deformations is rendered unstable. A scaling argument is
supplied that allows the calculation of the effective macroscopic exponents
starting from this microscopic picture.

It seems that the effects described above can only be accounted for by going
beyond the conventional approach that considers {\it typical polymer segments}
only.  Instead, we propose to describe the elasticity in terms of a {\it typical
  polymer filament} and the spatial distribution of cross-links along its
backbone. By controlling the architecture of the network, the scale of the
polymer length $l_f$ therefore seems to implicitly influence the elastic
properties of the system even in parameter regions where it does not enter the
macroscopic elastic moduli explicitly.

\begin{acknowledgments}
  We gratefully acknowledge fruitful discussions with Mark Bathe, Oskar
  Hallatscheck and Klaus Kroy.
\end{acknowledgments}

\appendix


  

\section{Stiffness Matrix}\label{sec:stiffness-matrix}

This appendix derives an expression for the stiffness matrix of a polymer
segment imbedded in a two-dimensional network.

The differential equation governing the bending of a beam of length $l$ is given
by $\kappa X^{(4)}=0$, where the transverse deflection $X$ is induced by the
forces $F_0$, $F_l$ as well as the torques $M_0$, $M_l$ acting on both ends. The
solution can then be written as
\begin{equation}\label{eq:bend}
  X(s) = X_0+X_0's+\frac{s^2}{2\kappa}(M_0-sF_0/3)\,,
\end{equation}
while equilibrium conditions require that
\begin{equation}\label{eq:staticBend}
  F_l=-F_0\,,\qquad M_l = -(M_0-F_0l)\,.
\end{equation}

Stretching the beam to the position $Z$ is governed by the equation
\begin{equation}\label{eq:stretch}
Z(s) = Z_0+s-\frac{s}{EA}T_0\,,
\end{equation}
with the condition
\begin{equation}\label{eq:staticStretch}
T_l = -T_0\,,
\end{equation}
balancing the axial forces $T$. 

The two variables $(X,Z)$ are the coordinates (in the frame of the fiber) of the
vector $\mathbf{u}$ introduced in the main text. The rotation is given by
$\theta=X'$. The four
Eqs.~(\ref{eq:bend}),~(\ref{eq:staticBend}),~(\ref{eq:stretch}) and
(\ref{eq:staticStretch}) can now be inverted to yield the forces in terms of the
displacements at the beam ends (cross-links)

\begin{widetext}
  \begin{equation}\label{eq:stiffMatrix}
    \left( \begin{array}[c]{c} F_0 \\ T_0 \\ M_0l \\ F_l \\ T_l \\ M_ll
      \end{array}\right)
    = \frac{\kappa}{l^3} \left(
      \begin{array}[c]{cccccc}
        -12 & 0   & -6 & 12 & 0 & -6 \\
        0   &   \Lambda & 0  & 0 & -\Lambda & 0 \\
        -6 & 0 & -4 & 6 & 0 & -2\\
        12 & 0 & 6  & -12 & 0 & 6  \\
        0   &  -\Lambda & 0  & 0 & \Lambda & 0 \\
        -6 & 0 & -2 & 6 & 0 & -4\\
      \end{array}\right)
    \cdot \left(
      \begin{array}[c]{c}
        X_0 \\ Z_0 \\ X_0'l \\ X_l \\ Z_l \\ X_l'l
      \end{array}\right)\,,
  \end{equation}
\end{widetext}
where we have defined $\Lambda = l^2A/I = 4 (l/r)^2$ The second
equality only holds for circular beam cross-sections, where the moment of area
$I=\pi r^4/4$.  The corresponding
matrix is called the stiffness matrix.

If, in addition to Eq.(\ref{eq:stretch}), we assume that the stretching response
is governed by that of a thermally fluctuating stiff polymer we have to take
into account $k_\parallel$ of Eq.(\ref{eq:k_entr}). This is achieved by letting
both stretching modes act in series and substitute $k_s^{-1} \to k_s^{-1} +
k_\parallel^{-1}$. Equivalently, one can assign an effective polymer radius
\begin{equation}\label{eq:polRadius}
r^2_{pol} = r^2 + \frac{4l^3}{\zeta l_p}\,,
\end{equation}
which now depends on the segment length $l$ as well as on the persistence length
$l_p$ of the polymer.


\end{document}